\documentclass[referee]{raa}
\usepackage{graphicx,times,subfigure}
\usepackage{natbib}
\usepackage{booktabs}
\usepackage{threeparttable}
\usepackage{amssymb,amsmath}
\usepackage{cases}
\bibpunct{(}{)}{;}{a}{}{,}

\usepackage[a4paper=true,dvipdfm=true,pagebackref=true]{hyperref}
\hypersetup{pdftitle = The title of my PDF, pdfauthor = My name, pdfsubject= The subject, pdfkeywords = keyword1 keyword2 keyword3}
\hypersetup{colorlinks = true, linkcolor = green, anchorcolor = red, citecolor = blue, filecolor = red, pagecolor = red, urlcolor = red}

\begin{document}

   \title{Using Chebyshev polynomials interpolation to improve the computation efficiency of gravity near an irregular-shaped asteroid
$^*$
\footnotetext{\small $*$ Supported by the National Natural Science Foundation of China.}
}

 \volnopage{ {\bf } Vol.\ {\bf X} No. {\bf XX}, 000--000}
   \setcounter{page}{1}

   \author{Shou-Cun Hu\inst{1,2} and Jiang-Hui Ji\inst{1}
   }

   \institute{\inst{1}CAS Key Laboratory of Planetary Sciences, Purple Mountain Observatory, Chinese Academy of Sciences, Nanjing 210008, China; \\
   \inst{2}University of Chinese Academy of Sciences, Beijing 100049, China;\\
   {\it jijh@pmo.ac.cn}\\
}

\abstract{In asteroid rendezvous missions, the dynamical environment near the asteroid's surface should be made clear prior to the  mission launch. However, most of the asteroids have irregular shapes, which lower the efficiency of calculating their gravitational field by adopting the traditional polyhedral method. In this work, we propose a method to partition the space near the asteroid adaptively along three spherical coordinates and use Chebyshev polynomials interpolation to represent the gravitational acceleration in each cell. Moreover, we compare four different interpolation schemes to obtain the best precision with the identical initial parameters. An error-adaptive octree division is combined to improve the interpolation precision near the surface. As an example, we take the typical irregular-shaped near-Earth asteroid 4179 Toutatis to show the advantage of this method, as a result, we show that the efficiency can be increased by hundreds to thousands times with our method. In a word, this method can be applicable to other irregular-shaped asteroids and can greatly improve the evaluation efficiency.
 }

   \authorrunning{S.-C. Hu et al. }            
   \titlerunning{Gravity near an irregular-shaped asteroid}  
   \maketitle

\section{Introduction}
Asteroids are thought to be leftover planetesimals related to the precursor bodies which formed the planets in our solar system. The primitive asteroids may provide a record of the original composition of the solar nebula where the planets are born. And the water and organic matter can provide us important clues on the origin of life on Earth. In additions, the near-Earth asteroids, whose orbits may cross the Earth orbit, may pose a potential risk to human beings on Earth\citep{michel2015asteroids}.

Through the ground- and space-based observations, the asteroid mission of flybys, rendezvous and landing, and laboratory analysis of return samples and all kinds of meteorites, we have made tremendous advances in the knowledge of asteroids \citep{nesvorny2015asteroid}. Among these techniques, space missions can directly show the detailed information in the closest distance. Since the first close-up image of asteroid 951 Gaspra taken in 1971 by Galileo spacecraft en route to Jupiter, 13 asteroids (including dwarf planet Ceres and Pluto) have been explored by spacecrafts. On December 13, 2012, the Chinese lunar probe Chang'e-2 flew by Toutatis with a surface distance of 0.77 km \citep{huang2013ginger, jiang2015boulders, zhao2015orientation}. Recently, OSIRIS-REx was launched by NASA on September 8, 2016 and now is on its way to asteroid 101955 Bennu \citep{lauretta2012overview}. Besides, the Hayabusa 2 mission, launched by JAXA in December 2014, will arrive at asteroid 162173 Ryugu in July 2018 \citep{muller2017hayabusa}. Both of the spacecrafts will bring the sample dust from the asteroids back to Earth.

The gravity field is essential to understand the dynamical environment of the asteroid, especially for the orbit design of spacecraft near asteroid (in orbiting phase or landing phase). The images captured by the spacecrafts truly revealed the fact that most asteroids have irregular shapes, different from planets approximate to spherical shape. The irregular shape of the asteroids causes the difficulty in calculating its gravity field. The former investigations show that three major approaches of spherical/ellipsoidal harmonics expansion, polyhedral method and mascon approximation based on finite element representation, have been developed to evaluate the gravity. Among them, spherical harmonics method is based on series expansion\citep{kaula1966theory, lundberg1988recursion, hu2015application}, which may not converge inside the so-called Brillouin sphere \citep{brillouin1933equations}. Though ellipsoidal harmonics expansion has larger convergence region \citep{romain2001ellipsoidal, garmier2002modeling}, the computation of ellipsoidal harmonics are not so straightforward and it does not fundamentally resolve the convergence problem. Recently, \cite{takahashi2014small} proposed to use interior spherical harmonic expansion to extend the convergence region within the interior Brillouin sphere. However, this method is not suitable to be practically used due to its complexity. Assuming a constant density, polyhedral method may be utilized to precisely evaluate the gravity field \citep{werner1996exterior}. Mascon approximation uses collection of cubes or spheres to represent the true internal structure of asteroids\citep{park2010estimating, chanut2015mascon, zhao2016using}. However, both of them are computationally intensive, and situation will get worse if the number of the facets and vertexes or mascons increases. This problem is particularly concerned for large-amount simulations (such as Monte Carlo analysis) or smaller onboard computation ability due to its relatively light load on the processor.

Several techniques were proposed to minimize the computation time in polyhedral method, such as using simpler approximations to the more computationally intensive terms in the formula, or adopting a coarser shape model at the expense of accuracy \citep{cangahuala2005augmentations, weeks2002gravity}. In this work, we introduce Chebyshev polynomials interpolation to accelerate the computation efficiency\citep{mason2002chebyshev}, which has been widely used in numerical representation of planetary ephemerides for years, such as the DE-series ephemeris developed by JPL and the INPOP ephemeris developed by France \citep{folkner2014planetary, fienga2008inpop06}. Actually, it was initially  put forward to speed up the calculation efficiency of Earth gravity by \cite{smith1981techniques}, in which Chebyshev expansions are applied only to the part of the gravity force expressed by spherical harmonic terms of degree larger than 4. However, the case for asteroids is quite different when considering the harmonics convergence problem above-mentioned. Herein we will refine this method to make it suitable to deal with irregular-shaped asteroids by applying new schemes.

In Section 2, we will firstly introduce our method in detail, including the space partition method, the comparison of the four interpolation schemes and the error-adaptive octree division. In Section 3, we will show the computation efficiency and orbit integration precision with numerical simulations, by comparing the results with those of polyhedral method. Finally, we present a brief
conclusion.

\section{Method}
In mathematics, the Chebyshev polynomials of the first kind are a sequence of orthogonal polynomials defined as the solutions of the Chebyshev differential equation \citep{rivlin1990chebyshev}. They may be calculated recursively as follows

\begin{equation}\label{eq:cheby0}
\left\{ \begin{array}{l}
{T_0}(x) = 1\\
{T_1}(x) = x\\
{T_{n + 1}}(x) = 2x{T_n}(x) - {T_{n - 1}}(x)
\end{array} \right.
\end{equation}
where the range of x is $-1 \leq x \leq 1$. Chebyshev polynomials are stable during evaluation, and they provide a readily apparent estimate of neglected terms on interpolation error. Besides the high computational efficiency, the resulting interpolation polynomial also minimize the Runge's phenomenon problem and provide an approximation that is close to the polynomial of best approximation to a continuous function under the maximum norm \citep{hernandez2001chebyshev}.

As mentioned above,  Chebyshev polynomials are widely used in numerical representation of planetary ephemerides \citep{newhall1989numerical}. During the process, the range of time is segmented into contiguous intervals of fixed length and then the interpolation of rectangular coordinates is performed in each segment. Back to gravitational acceleration near an asteroid, \textbf{it is similar to} represent it as Chebyshev polynomials, except that we should consider three-dimensional Chebyshev polynomials interpolation in this situation. The basic formula is
\begin{equation}\label{eq:fitting0}
{\bf{F}}\left( {r,\theta ,\varphi } \right) = \left( {\begin{array}{*{20}{l}}
{\sum\limits_{i = 0}^N {\left[ {\sum\limits_{j = 0}^N {\left( {\sum\limits_{k = 0}^N {C_{ijk}^{(1)}{T_k}\left( {\tilde \varphi } \right)} } \right){T_j}\left( {\tilde \theta } \right)} } \right]{T_i}\left( {\tilde r} \right)} }\\
{\sum\limits_{i = 0}^N {\left[ {\sum\limits_{j = 0}^N {\left( {\sum\limits_{k = 0}^N {C_{ijk}^{(2)}{T_k}\left( {\tilde \varphi } \right)} } \right){T_j}\left( {\tilde \theta } \right)} } \right]{T_i}\left( {\tilde r} \right)} }\\
{\sum\limits_{i = 0}^N {\left[ {\sum\limits_{j = 0}^N {\left( {\sum\limits_{k = 0}^N {C_{ijk}^{(3)}{T_k}\left( {\tilde \varphi } \right)} } \right){T_j}\left( {\tilde \theta } \right)} } \right]{T_i}\left( {\tilde r} \right)} }
\end{array}} \right)
\end{equation}
\textbf{where $r$, $\theta$ and $\varphi$ are the three spherical coordinates in body-fixed reference system, i.e. radial distance, longitude and latitude, respectively. $\tilde r,\tilde \varphi $ and $\tilde \theta$ are defined as
\begin{equation}\label{eq:bianhua}
\tilde r = \frac{{2r - {r_{\max }} - {r_{\min }}}}{{{r_{\max }} - {r_{\min }}}},\;\;\tilde \varphi  = \frac{{2\varphi  - {\varphi _{\max }} - {\varphi _{\min }}}}{{{\varphi _{\max }} - {\varphi _{\min }}}},\;\;\tilde \theta  = \frac{{2\theta  - {\theta _{\max }} - {\theta _{\min }}}}{{{\theta _{\max }} - {\theta _{\min }}}}
\end{equation}
where $r_{\min}, r_{\max}, \varphi _{\min}, \varphi _{\max}, \theta _{\min}$ and $\theta _{\max}$ are minimal or maximal value of $r, \varphi$ and $\theta$ of the domain. $T_k$ are the Chebyshev polynomials defined in \eqref{eq:cheby0}}. $\bf{F}$ is gravitational acceleration vector and $C_{ijk}^{(1)}$, $C_{ijk}^{(2)}$, $C_{ijk}^{(3)}$ are Chebyshev polynomials coefficients of each component with degree $N$ (we have assumed the same degree in three components), \textbf{which may be solved by least-squares method}.

\subsection{Division scheme}
\begin{figure}
\includegraphics[width=1.00\textwidth]{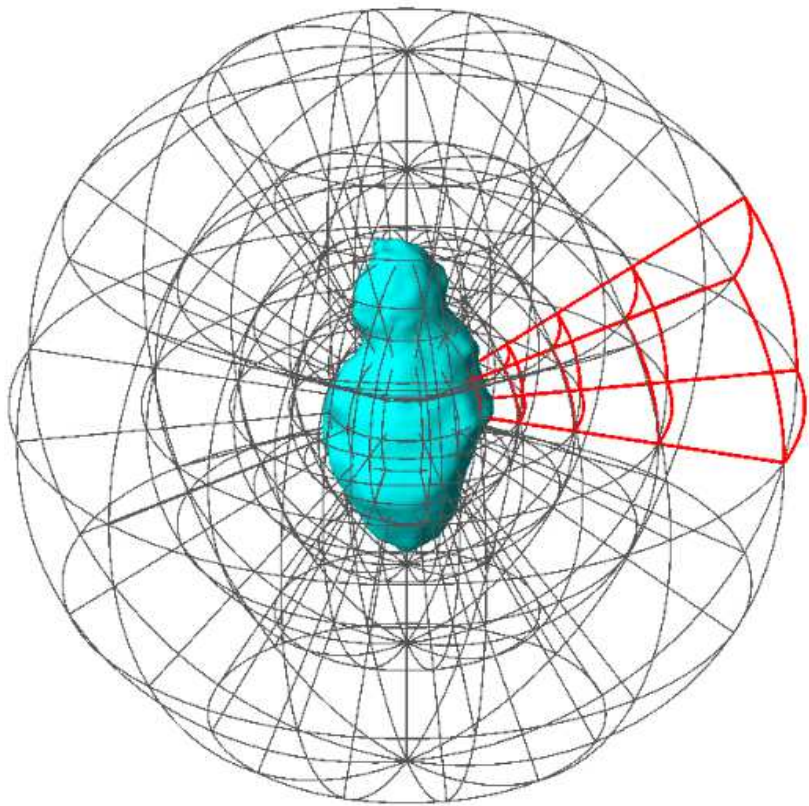}
\caption{The illustration of division of the neighborhood space along the spherical coordinates directions, taking asteroid 4179 Toutatis for example. The red outlines show the division along some radial direction. The shape model is credited by \cite{hudson2003high}.}
\label{fig:Toutatis_divide}
\end{figure}

In our method, the space near asteroid is divided along $r, \theta$ and $\varphi$ (we call it spherical division scheme hereafter). Asteroid 4179 Toutatis is a typical irregular-shaped asteroid, with dimension x = 4.60 km, y = 2.29 km and z = 1.92 km \citep{hudson2003high, huang2013ginger}. Take Toutatis as example, the division is illustrated in Fig. \eqref{fig:Toutatis_divide}, where the range of each coordinate in each cell is represented as

\begin{equation}\label{eq:cell_range}
\left\{ \begin{array}{l}
\Delta {r_i} = r_{\max }^i - r_{\min }^i\\
\Delta {\theta _i} = \theta _{\max }^i - \theta _{\min }^i\\
\Delta {\varphi _i} = \varphi _{\max }^i - \varphi _{\min }^i
\end{array} \right.
\end{equation}

In the illustration, the asteroid is divided uniformly along longitude and latitude direction (let $\Delta {\theta _i} = \Delta {\varphi _i} = \alpha$). However, in radial direction the range is picked so that it is nearly proportional to radial distance, i.e.
\begin{equation}\label{eq:delta_r}
\Delta {r_i} = r_{\min }^i\sin \alpha
\end{equation}
The trick above is based on the fact that the variation of gravitational acceleration is gentler at further distance, so we can use larger $\Delta {r_i}$ for larger $r$, so as to reduce the storage amount of coefficients. The error of interpolation along radial direction will be shown hereinafter. In this scheme, we can use $\alpha$ and $N$ to adjust the precision (both smaller $\alpha$ and larger $N$ may reduce the error, but demand for larger storage), and $r_{min}, r_{max}$ to constrain the domain we are interested in. Of course, $r_{min}$ is usually chosen as the minimal radial distance at the surface.

In programming, we only need to load the coefficients once, and then the computation time of $\bf{F(r)}$ almost only depends on $N$ \textbf{(as we can see in \eqref{eq:fitting0}, the calculation is not related with $\alpha$)}. The whole procedure includes coefficients generation and gravity calculation. Polyhedral method is used during the process of coefficients generation.

\subsection{Comparison of four interpolation schemes}
As we know, gravitational acceleration can be divided into central part and non-spherical part. \textbf{Thus Eq. \eqref{eq:fitting0} is modified as \citep{kaula1966theory}}

\begin{equation}\label{eq:fitting1}
{\bf{F}}\left( {r,\theta ,\varphi } \right) = {{\bf{F}}_0}\left( {r,\theta ,\varphi } \right) + K\left( r \right) \cdot {\bf{F}'}(r,\theta ,\varphi )
\end{equation}

\begin{equation}\label{eq:fitting2}
{\bf{F}}'\left( {r,\theta ,\varphi } \right) = \left( {\begin{array}{*{20}{l}}
{\sum\limits_{i = 0}^N {\left[ {\sum\limits_{j = 0}^N {\left( {\sum\limits_{k = 0}^N {C_{ijk}^{(1)}{T_k}\left( {\tilde \varphi } \right)} } \right){T_j}\left( {\tilde \theta } \right)} } \right]{T_i}\left( {\tilde r} \right)} }\\
{\sum\limits_{i = 0}^N {\left[ {\sum\limits_{j = 0}^N {\left( {\sum\limits_{k = 0}^N {C_{ijk}^{(2)}{T_k}\left( {\tilde \varphi } \right)} } \right){T_j}\left( {\tilde \theta } \right)} } \right]{T_i}\left( {\tilde r} \right)} }\\
{\sum\limits_{i = 0}^N {\left[ {\sum\limits_{j = 0}^N {\left( {\sum\limits_{k = 0}^N {C_{ijk}^{(3)}{T_k}\left( {\tilde \varphi } \right)} } \right){T_j}\left( {\tilde \theta } \right)} } \right]{T_i}\left( {\tilde r} \right)} }
\end{array}} \right)
\end{equation}
where ${\bf{F}}_0$ is the part that can be calculated analytically and $K(r)$ is a scalar coefficient related to $r$. Let's consider the three different schemes (denoted as I1, I2 and I3)

\begin{equation}\label{eq:fitting3}
\left\{ \begin{array}{l}
{\rm{I}}1:\;\;\;{{\bf{F}}_0}\left( {r,\theta ,\varphi } \right) = 0,\;\;\;\;\;\;\;\;\;\;\;\;\;K(r) = 1\\
{\rm{I}}2:\;\;\;{{\bf{F}}_0}\left( {r,\theta ,\varphi } \right) =  - \frac{{GM}}{{{r^3}}}{\bf{r}},\;\;\;\;K(r) = 1\\
{\rm{I}}3:\;\;\;{{\bf{F}}_0}\left( {r,\theta ,\varphi } \right) =  - \frac{{GM}}{{{r^3}}}{\bf{r}},\;\;\;\;K(r) = \frac{{GM}}{{{r^4}}}
\end{array} \right.
\end{equation}
where I1 is the same as Eq.\eqref{eq:fitting0}, but I2 and I3 only fit non-spherical part. I3 also considers the characteristics that the generally largest zonal and tesseral harmonics gravitational acceleration ${\bf{F}}_{J_2}$ and ${\bf{F}}_{J_{22}}$ \textbf{are inversely proportional to the fourth power of $r$ \citep{kaula1966theory}}

\begin{equation}\label{eq:J2_J22}
{{\bf{F}}_{{J_2}}}\sim\frac{{GM}}{{{r^4}}},\;\;\;\;{{\bf{F}}_{{J_{22}}}}\sim\frac{{GM}}{{{r^4}}}
\end{equation}

Besides, we may consider another interpolation scheme. In Eq. \eqref{eq:fitting1}, \eqref{eq:fitting2} and \eqref{eq:fitting3}, the gravitational acceleration is represented with rectangular components ${F_x},{F_y}$ and ${F_z}$ by default. However, we can also try to transform them to three components ${F_r},{F_{\theta}}$ and ${F_{\varphi}}$ along the spherical coordinates, i.e.

\begin{equation}\label{eq:transformation}
\left( {\begin{array}{*{20}{c}}
{{F_r}}\\
{{F_\theta }}\\
{{F_\varphi }}
\end{array}} \right) = {\bf{A}}\left( {\begin{array}{*{20}{c}}
{{F_x}}\\
{{F_y}}\\
{{F_z}}
\end{array}} \right)
\end{equation}
Here $\bf{A}$ is the transformation matrix
\begin{equation}\label{eq:matrix_A}
{\bf{A}} = {{\bf{R}}_x}(\frac{1}{2}\pi  - \varphi ){{\bf{R}}_z}(\theta  - \frac{3}{2}\pi )
\end{equation}
in which ${{\bf{R}}_x(\theta)}$ and  ${{\bf{R}}_z(\theta)}$ are defined as

\begin{equation}\label{eq:RxRy}
{{\bf{R}}_x}(\theta ) = \left( {\begin{array}{*{20}{c}}
1&0&0\\
0&{\cos \theta }&{\sin \theta }\\
0&{ - \sin \theta }&{\cos \theta }
\end{array}} \right),\;\;{{\bf{R}}_z}(\theta ) = \left( {\begin{array}{*{20}{c}}
{\cos \theta }&{\sin \theta }&0\\
{ - \sin \theta }&{\cos \theta }&0\\
0&0&1
\end{array}} \right)
\end{equation}
Besides, we use the same value of scheme I3 for ${{\bf{F}}_0}\left( {r,\theta ,\varphi } \right)$ and $K(r)$. This is denoted as I4 scheme.

\begin{figure}
\includegraphics[width=1.00\textwidth]{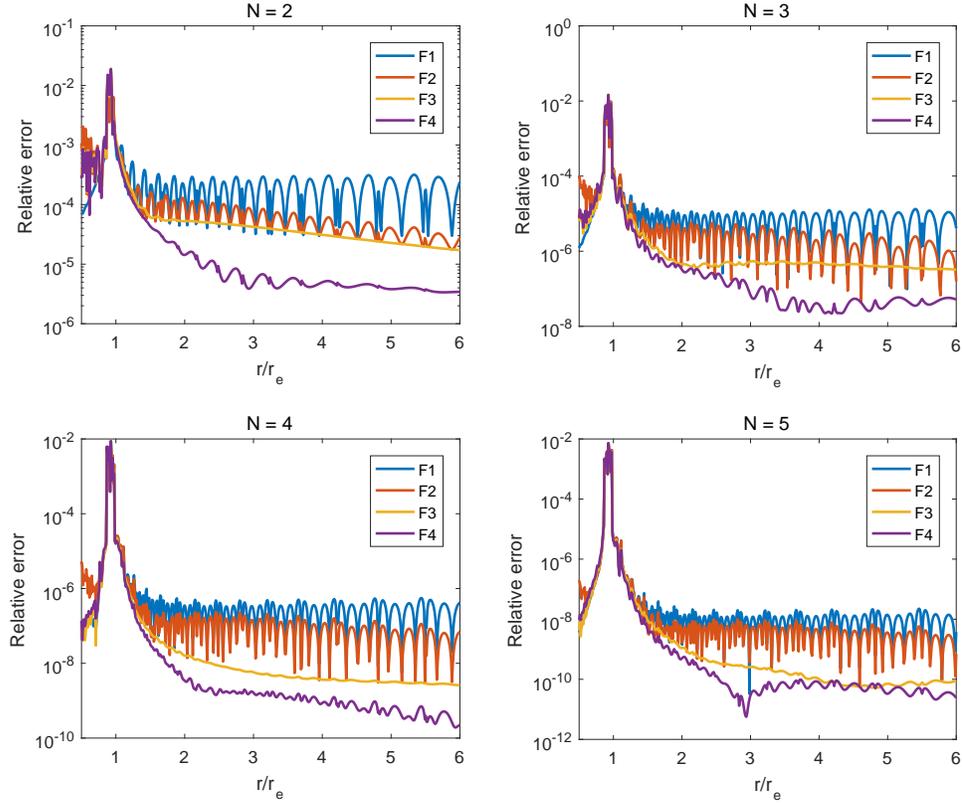}
\caption{The relative error vs. $r$ for four interpolation schemes I1, I2, I3 and I4 at direction $\theta=0^\circ$ and $\varphi=0^\circ$. The degrees of Chebyshev polynomials are 2, 3, 4 and 5, respectively. $r_e$ is radial distance on the surface.}
\label{fig:F_vs_r}
\end{figure}

Now, let's compare the interpolation results for Toutatis. Set $\alpha  = 10^\circ$, $N = 2,3,4,5$. The relative error $\delta_{\bf{F}}$ is defined as
\begin{equation}\label{eq:rel_err}
\delta_{\bf{F}} = \frac{{|{\bf{F}} - {\bf{\tilde F}}|}}{{|{\bf{F}}|}}
\end{equation}
where $\bf{F}$ is the gravitational acceleration calculated with polyhedral method \textbf{(see Eq.15 in \citep{werner1996exterior})} and $\bf{\tilde F}$ is the value by interpolation. The comparisons of $\delta_{\bf{F}}$ for the four interpolation schemes are shown in Fig. \eqref{fig:F_vs_r}, where $\delta_{\bf{F}}$ varies with $r$. In the results, relative errors reach maximum near the surface, which is because of the violent gravity change in this area. Anyway, we can see that I1 has the best result inside the asteroid. But outside, I3 is better than I1 and I2. While I4, plus a transformation on I3, has a better result than I3, especially for lower degree $N$. Because we only care about the gravity outside the asteroid for the most situations, here we recommend using I4 scheme in our method.

\subsection{Comparison between spherical and rectangular division schemes}
\begin{figure}
\includegraphics[width=1.00\textwidth]{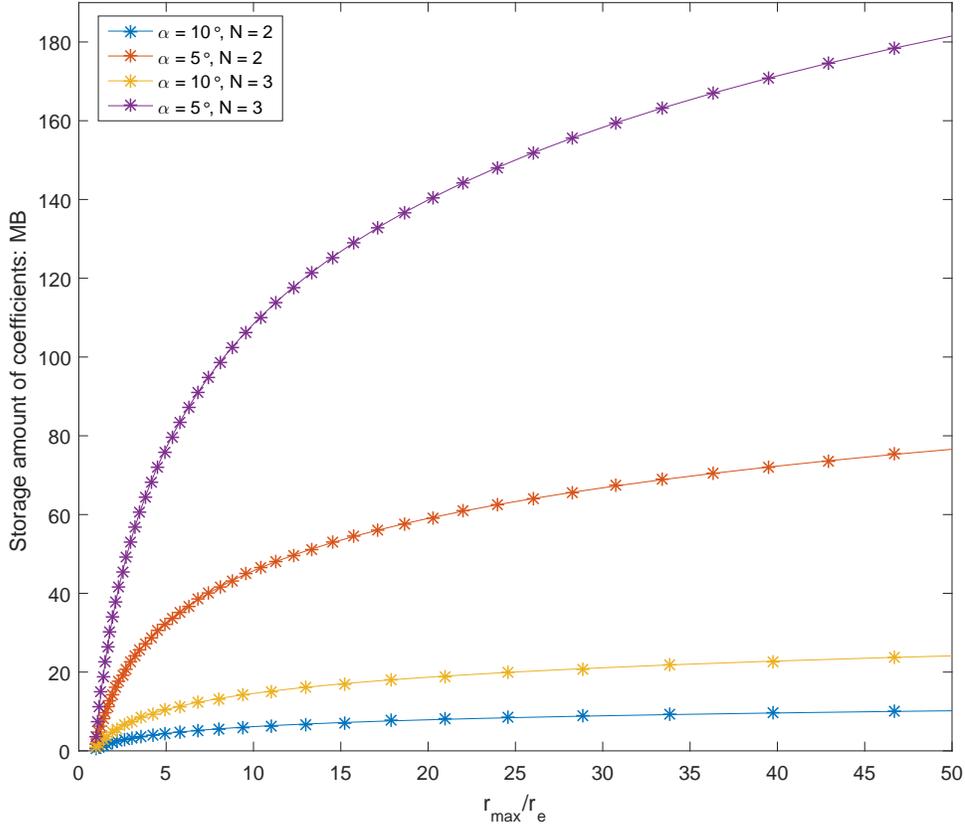}
\caption{The storage amount of Chebyshev coefficients vs. $r_{max}$ ($r_{min} = r_e$) (Double-precision binary floating-point is assumed for the storage of coefficients).}
\label{fig:storage}
\end{figure}

As Fig. \eqref{fig:F_vs_r} shows, for all of the four schemes, the relative errors generally do not increase with $r$ when $r/r_e > 1$, by which we show that the linear increase of radial range of each cell in Eq.\eqref{eq:delta_r} is a reasonable choice for our spherical division scheme. As mentioned above, this will reduce the storage amount for large $r_{max}$. Actually, we may estimate the storage amount of Chebyshev coefficients for a specific $\alpha$ and $N$. As illustrated in Fig. \eqref{fig:storage}, the speed of \textbf{storage increment} with $r_{max}$ is decreasing as $r$ increases. That means we do not need to worry about the storage of coefficients too much when we need to consider a large $r_{max}$. Besides, as we can see in Fig.\eqref{fig:F_vs_r}, the error is satisfactory at large $r$. So we may not need to replace our method by switching to use spherical harmonic expansion to calculate the gravity at large $r$.

Despite the advantage above, we are still interested in the comparison between spherical division and rectangular division. For the latter, the division is along the directions of three rectangular coordinates x, y and z, in which the cells are actually cubes (assume $\Delta {x_i} = \Delta {y_i} = \Delta {z_i} = D$ for all the cells) in rectangular coordinate system.

\begin{figure}
\includegraphics[width=1.00\textwidth]{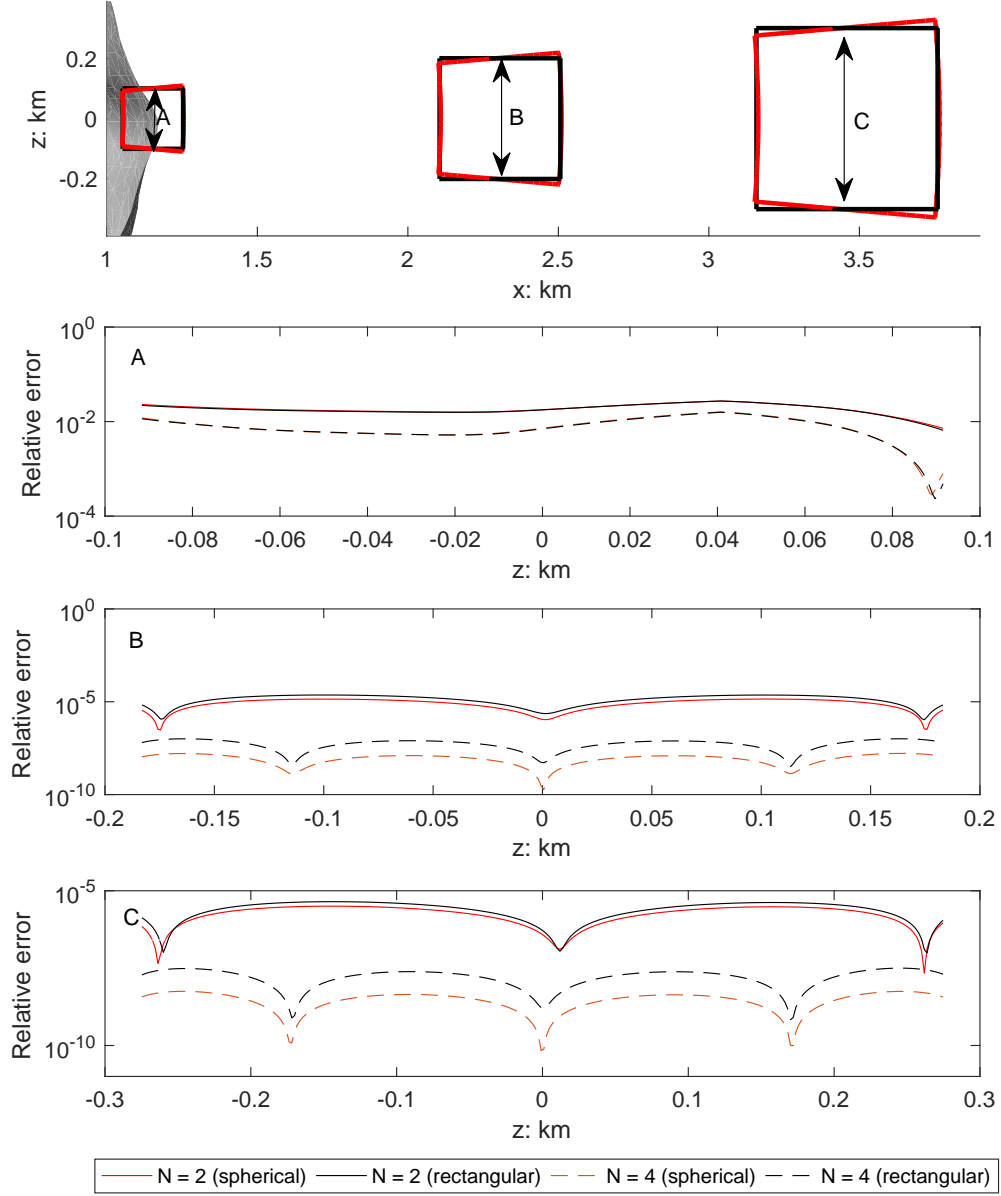}
\caption{The comparison of relative error for spherical and rectangular division schemes in area A, B and C. The I4 interpolation scheme is adopted in the calculation.}
\label{fig:ABC}
\end{figure}

Consider three different cells near Toutatis, which are divided in a spherical way with $\alpha = 10^\circ$, as illustrated in the top panel of Fig. \eqref{fig:ABC}, in which A, B and C is the center of each cell and $r_C = \frac{3}{2} r_B = 3 r_A$. Only cell A crosses the interior and exterior part of Toutatis. The red outline is the profile of the three cells projected in plane y = 0. We can also have a rectangular division at A, B and C, with the length of each cube equaling to the average scale of cell A, B and C, respectively. They are illustrated as the black outline in the panel.

The relative errors along z-axis (the position along the double-arrows) crossing each center of A, B and C are given in Fig. \eqref{fig:ABC}, where $N = 2, 4$ and both spherical and rectangular division results are shown. We can see that the two division schemes almost do not show any difference in area A, but spherical division prevails in area B and C (For B area, the mean relative error ratios of  spherical to rectangular division are 0.6 and 0.16 for $N = 2$ and $N = 4$, respectively, and For C area, the values are 0.72 and 0.18). This experiment concludes that spherical division scheme has an advantage over rectangular division far from the surface of asteroid, this is another reason we recommend using spherical division.

\subsection{Combined with octree division}
\begin{figure}
\includegraphics[width=1.00\textwidth]{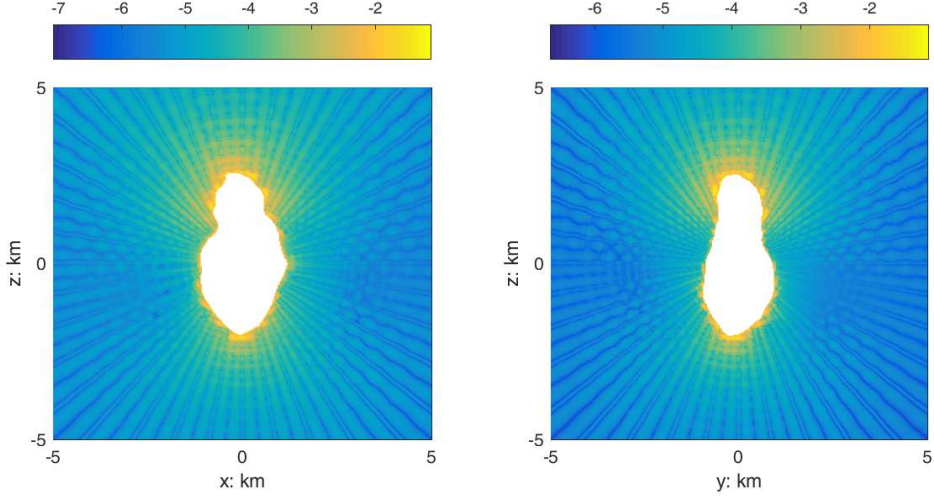}
\caption{The profiles of relative errors at $y = 0$ (left) and $x = 0$ (right) near asteroid Toutatis, where $N = 2$, $\alpha = 10^\circ$. We have applied $\rm log_{10}$ operation on the relative error. The interior gravity is not calculated.}
\label{fig:poumian1}
\end{figure}

We have done the numerical experiments to show that I4 interpolation and spherical division scheme are good choices when we use Chebyshev polynomials to approximate the gravity near an irregular-shaped asteroid. Set $ N = 2$ and $\alpha = 10^\circ$, the profile of relative errors of Toutatis in plane $y = 0$ and $x = 0$ are shown in Fig. \eqref{fig:poumian1}. We can see that the relative error may increase to $\sim$0.1 near the surface and decrease to less than 0.001 for $r > 3\;\rm km$, which is consistent with the results given in Fig. \eqref{fig:F_vs_r}. To improve the precision, we may reduce the size of each cell by decreasing $\alpha$ or increase the degree $N$. The experiments are performed for area A, B and C of Fig. \eqref{fig:ABC}. Fix $N = 2$ and halve $\alpha$ from $20^\circ$ to $1.25^\circ$. Fix $\alpha = 10^\circ$ and increase $N$ from 2 to 6. Their results of maximal relative errors are shown in Tab. \eqref{tab:wucha1} and \eqref{tab:wucha2}, respectively.

\begin{table}[htbp]
\centering
\caption{The maximal relative errors of cell A, B and C. $N = 2$, and $\alpha$ varies from $20^\circ$ to $1.25^\circ$}
\label{tab:wucha1}
\begin{tabular}{lccccc}
\toprule  & $\alpha = 20^\circ$ & $\alpha = 10^\circ$ & $\alpha = 5^\circ$ & $\alpha = 2.5^\circ$ & $\alpha = 1.25^\circ$ \\
\midrule
A & 1.09E-01 & 5.20E-02 & 2.23E-02 & 1.47E-02 & 8.82E-03 \\
B & 6.36E-04 & 5.55E-05 & 5.84E-06 & 6.72E-07 & 8.06E-08 \\
C & 1.89E-04 & 1.92E-05 & 2.18E-06 & 2.61E-07 & 3.21E-08 \\
\bottomrule
\end{tabular}
\end{table}

\begin{table}[htbp]
\centering
\caption{The maximal relative errors of cell A, B and C. $\alpha = 10^\circ$, and increase $N$ from 2 to 6}
\label{tab:wucha2}
\begin{tabular}{lccccc}
\toprule & $N = 2$ & $N = 3$ & $N = 4$ & $N = 5$ & $N = 6$ \\
\midrule
A & 1.13E-02 & 1.11E-02 & 6.36E-03 & 6.12E-03 & 3.91E-03 \\
B & 2.92E-06 & 7.63E-08 & 6.21E-10 & 2.70E-11 & 3.58E-13 \\
C & 1.09E-06 & 2.55E-08 & 1.51E-10 & 2.42E-12 & 4.99E-14 \\
\bottomrule
\end{tabular}
\end{table}

\begin{figure}
\includegraphics[width=1.00\textwidth]{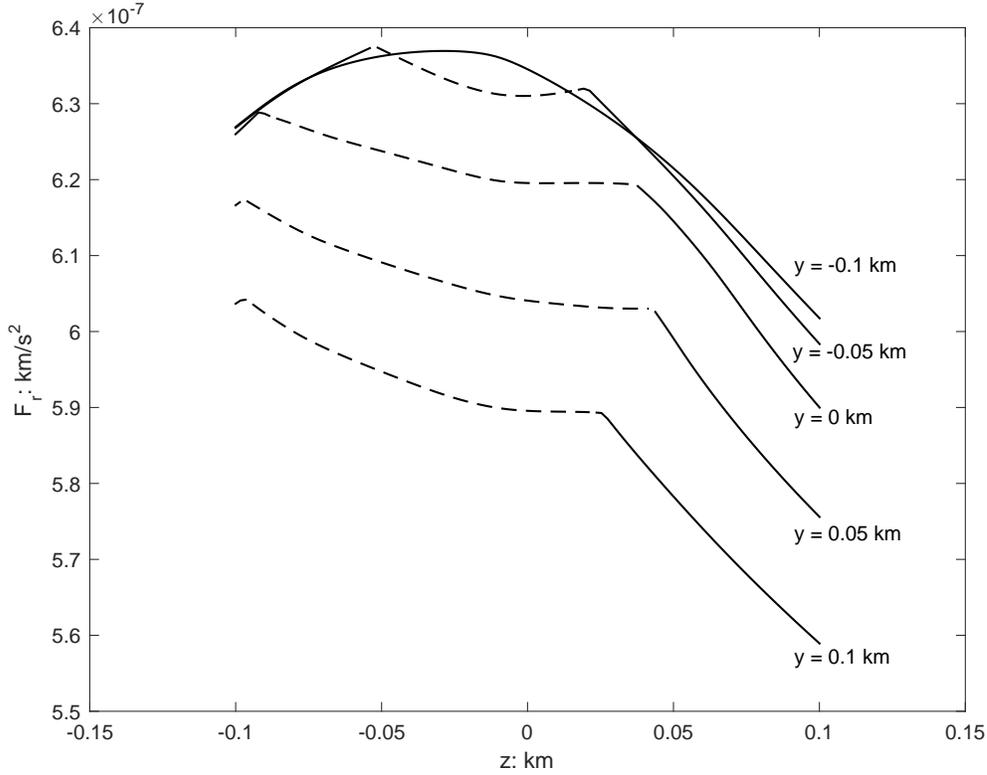}
\caption{$F_r$ vs. $z$ in plane $x= x_A$ for different $y$. The solid and dash lines represent the points are outside or inside the asteroid Toutatis, respectively.}
\label{fig:Fr_at_A}
\end{figure}

The results show that for each halving of $\alpha$, the maximal errors of B and C are reduced by 8$\sim$10 times, while the value is up to 50 times for each increase of $N$ from 2 to 6 on average. But the situation is totally different for A, where the errors are only reduced by less than 0.5 times for each halving of $\alpha$, and the situation is even worse for the increase of $N$. For an illustration, the variation of $F_r$ with $z$ in plane $x= x_A$ for different $y$ are shown in Fig. \eqref{fig:Fr_at_A}, where the dash lines mean the locations are inside the asteroid. The variations of $F_r$ with $z$ are continuous curves, but obvious jumps occur at the adjacent area. So it is a natural result for area A when we use smooth curve (Chebyshev polynomials) to fit the not-so-smooth gravity. If we need to refine the precision near the surface, a natural choice is to only reduce the cell size in these areas (we do not choose to increase the degree $N$ because this will cause the computation time enhancement, as we can see in Tab. \eqref{tab:time}), but not reduce $\alpha$ globally, because the latter one will unnecessarily reduce the cell size far away from the surface and greatly increase the storage burden (a rise of 8 times of storage for each halving of $\alpha$).

\begin{figure}
\includegraphics[width=1.00\textwidth]{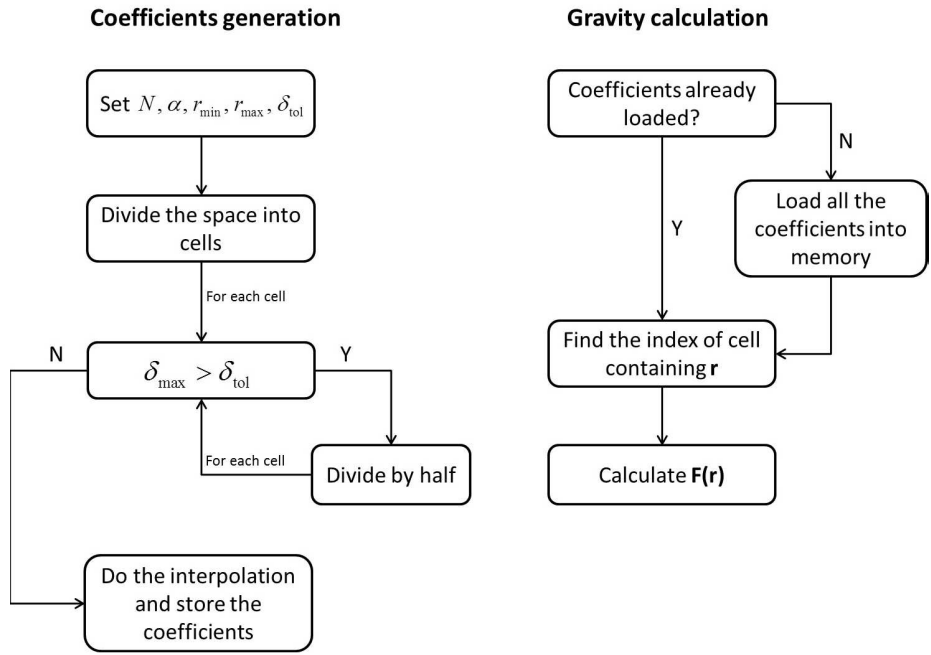}
\caption{The basic programming procedure of coefficients generation and gravity calculation. Polyhedral method is used in coefficients generation.}
\label{fig:liucheng2}
\end{figure}

With  the results above in mind, the spherical division scheme is refined by combining with adaptive octree division \citep{frisken2002simple}, in which a tolerant error (denoted as $\delta_{tol}$) is set in addition to $\alpha, N, r_{min}$ and $r_{max}$. The maximal relative error ($\delta_{max}$) of each cell is evaluated during the division. If $\delta_{tol} < \delta_{max}$, then the cell will be divided by half and this process is repeated recursively. Finally the cells may have different size and they are managed by an octree data structure. The whole procedure of coefficients generation and gravity calculation of our method is illustrated in Fig. \eqref{fig:liucheng2}.

\begin{figure}
\includegraphics[width=1.00\textwidth]{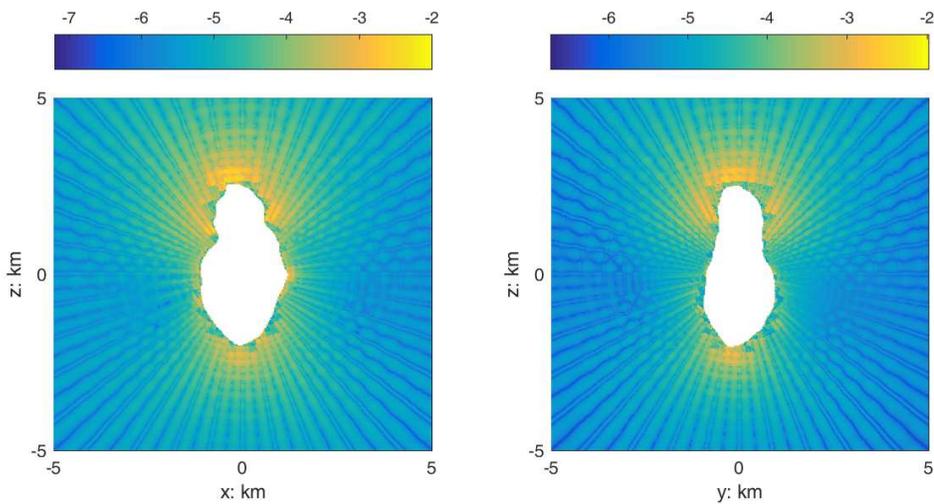}
\caption{The profiles of relative errors at $y = 0$ (left) and $x = 0$ (right) near asteroid Toutatis, where $N = 2$, $\alpha = 10^\circ$. The interior gravity is not calculated.}
\label{fig:poumian2}
\end{figure}

For an experiment, set $\delta_{tol} = 0.01$, and then Fig. \eqref{fig:poumian1} is refined as Fig. \eqref{fig:poumian2}, where the maximal relative errors near the surface have been reduced to less than 0.01.

\section{Computation efficiency and precision of orbit integration}
Based on the procedures in Fig. \eqref{fig:liucheng2}, we are able to generate the coefficients for asteroid Toutatis. Two important aspects of this method we are concerned are computation efficiency and error of orbit propagation.
\subsection{Comparison of computation efficiency with polyhedral method}
In our computation, the number of facets and vertexes of Toutatis' shape model are 6400 and 12796 \citep{hudson2003high}. After loading all the coefficients into memory, the computation efficiency almost only depends on degree $N$. Compared with polyhedral method, the elapsed time for $N = 2 \sim 6$ is list in Tab. \eqref{tab:time}.
\begin{table}[htbp]
\centering
\caption{The ratio of time consumption of our method to polyhedral method ($\alpha = 10^\circ$ and $N = 2 \sim 6$).}
\label{tab:time}
\begin{tabular}{lcccccc}
\toprule Method & Polyhedral & $N = 2$ & $N = 3$ & $N = 4$ & $N = 5$ & $N = 6$ \\
\midrule
Time consumption &  1.0 & 7.6E-4 & 1.2E-3 & 1.7E-3 & 2.4E-3 & 3.2E-3\\
\bottomrule
\end{tabular}
\end{table}

The results tell us that the computation efficiency of our method is hundreds to thousands times higher than polyhedral method, and this advantage is more prominent for shape models with larger number of facets and vertexes. As $N$ increases, the computing time also increases slightly. So in the sense of efficiency, for the same precision, we recommend using small $N$ and small $\alpha$, but not large $N$ and large $\alpha$.

\subsection{Errors comparison of orbit propagation with polyhedral method}
In most situations, the computation of gravitational acceleration is used in orbit integration. It is interesting to compare the orbit integration error between our method and polyhedral method.

\begin{figure}
\includegraphics[width=1.00\textwidth]{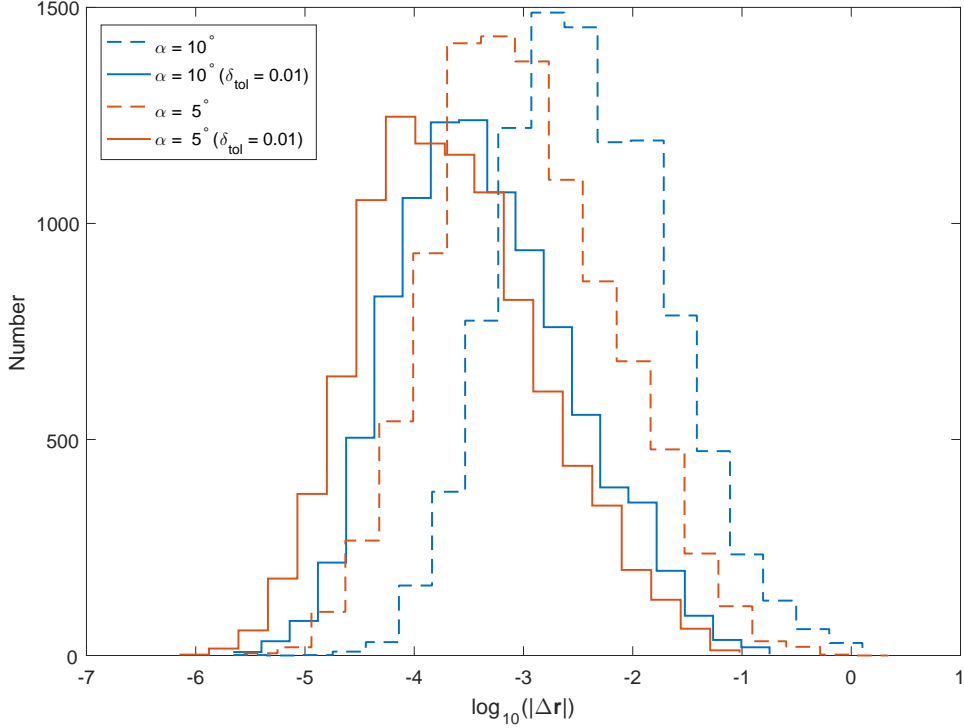}
\caption{The distribution of orbit error for ejecta particles. $N = 2$, $\alpha = 10^\circ$ or $5^\circ$. The dash lines do not set the value of $\delta_{tol}$ while solid lines have $\delta_{tol} = 0.01.$}
\label{fig:stairs_1}
\end{figure}

The gravity near the surface of asteroid is essential for proximation operations (includes hovering, landing and touch-and-go maneuvers) of spacecrafts. To show the application of our method in this circumstance, we have integrated 10, 000 orbits of ejecta particles randomly launching from the surface of Toutatis with launch angle of $45^{\circ}$ and velocities of $0.4 \sim 1.2$ m/s (the averaged escape velocity of Toutatis on the surface is $\sim$ 1.3 m/s) by using our method and polyhedral method to calculate the gravitational acceleration. Only the particles re-impact on the surface are recorded. About 5$\%$ particles have orbit time larger than 1 day, whose eccentricities are so large that their orbit errors are very sensitive to the gravity. So these particles are rejected because they are not suitable for comparison. For the remaining 95\% particles, the stairstep graph of position error distribution is given in Fig. \eqref{fig:stairs_1}, where $N = 2$, $\alpha = 10^\circ$ or $5^\circ$. The dash lines mean $\delta_{tol}$ is unset while solid lines have $\delta_{tol} = 0.01$. We can see that, the precision has an obvious enhancement after the combining with adaptive octree division, in which about 94.9 \% and 97.6 \% particles have errors less than 0.01 km for $\alpha = 10^\circ$ and $\alpha = 5^\circ$, respectively.

\begin{figure}
\includegraphics[width=1.00\textwidth]{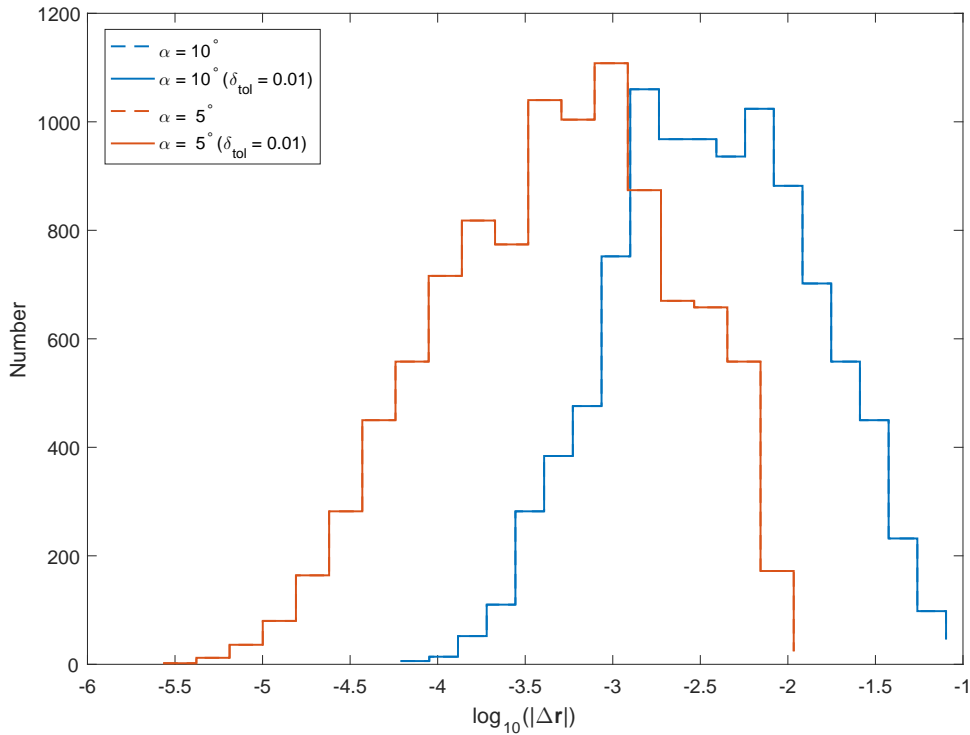}
\caption{The distribution of orbit error for circle orbits with $r$ = 4 km. The other parameters are the same with Fig. \eqref{fig:stairs_1}.}
\label{fig:stairs_2}
\end{figure}

Another situation we concern is the stable motion around the asteroid. In most cases, these orbits are high enough that we can safely use harmonic expansions to calculate the gravity. But this will bring additional trouble about harmonic coefficients retrieval. Nevertheless, here we would like to perform the experiments only using our method. 10, 000 particles are placed in circle orbits with $r$ = 4 km (about 2 times of the asteroid radius) and randomly picked inclinations and mean anomalies. The propagation is performed for 10 circles (about 5.6 days). Using the same parameters in Fig. \eqref{fig:stairs_1}, the results are shown in Fig. \eqref{fig:stairs_2}. This time, the dash lines and solid lines coincide because the adaptive octree division of $\delta_{tol} = 0.01$ does not influence the domains where these orbits pass through. About 99.9 \% and 79.1 \% orbits have errors less than 0.01 km for $\alpha = 5^\circ$ and $\alpha = 10^\circ$, respectively.

These two experiments conclude that, decreasing the cell size may definitely reduce the orbit error. For the orbits near the surface, using adaptive octree division is necessary and we can see it works well for the orbit integration. The orbit errors for $N = 2, \alpha = 5^\circ, \delta_{tol} = 0.01$ are acceptable in some situations, such as Monte Carlo simulation, or preliminary orbit design. But if you still need a higher precision, reducing $\delta_{tol}$ may be a good choice. However, please keep in mind that the rate of error reduction near the surface is very small as $\alpha$ decreases, which means you probably need a very high cost of coefficients storage to trade a little precision enhancement.

Finally, for a reference about the storage, if $r_{min}$ = 0.75 km, $r_{max}$ = 20 km, and ignore the cell totally inside Toutatis, the double-precision binary storage of coefficients is 54.7 MB and 7.4 MB for $\alpha = 5^\circ$ and $10^\circ$ with $\delta_{tol}$ unset, respectively, while the values are 107.4 MB and 61.6 MB with $\delta_{tol} = 0.01$, respectively.

\section{Conclusions}
In this article, we propose to use Chebyshev polynomials interpolation to increase the computation efficiency of gravitational acceleration near an irregular-shaped asteroid, in which the gravity of the neighbourhood domain of the asteroid is precomputed by computationally expensive polyhedral method and the interpolation coefficients are stored. Spherical division and rectangular division scheme, four interpolation schemes on different components of gravitational acceleration are both compared, and we recommend adopting spherical division and I4 interpolation scheme according to the numerical experiments performed on asteroid 4179 Toutatis. The spherical division we propose along the radial direction is not uniform, where $\Delta r$ of each cell is nearly proportional to the radial distance. It allows us to use our method to calculate the gravity globally for some orbits not too far away from the surface at the cost of not too much additional \textbf{storage increment}. The I4 interpolation scheme suggests represent the gravitational acceleration along the three spherical coordinates directions, and we only need to do the interpolation on non-spherical part with an extra consideration about the variation characteristic along radial direction.

After that we show the computational efficiency may have an enhancement to hundreds to thousands times for the typical asteroid Toutatis and the speed enhancement mainly depends on degree of polynomials. The orbit propagation experiments are performed for 10, 000 ejecta orbits and stable midrange orbits. The results tell us that we can obtain a generally acceptable orbit precision by simply setting the parameters $N = 2, \alpha = 5^\circ$ and $\delta_{tol} = 0.01$, and the storage amount of coefficients is also acceptable.

However, we also notice that there is an obvious balance between precision and storage amount of coefficients. And special concern is noted about the slow error convergence near the asteroid surface, this is a drawback about the Chebyshev polynomials interpolation, for which the violent gravity change near the surface greatly increases the interpolation error. The subsequent improved research should focus on this issue.

\begin{acknowledgements}
This work is financially supported by the National Natural Science Foundation of China (Grants No. 11473073, 11503091, 11661161013, 11633009), and Foundation of Minor Planets of the Purple Mountain Observatory.
\end{acknowledgements}

\end{document}